\documentclass[preprint,12pt]{elsarticle}

\parskip\medskipamount

\usepackage[latin1]{inputenc}
\usepackage[english]{babel}
\usepackage[bookmarks,a4paper,colorlinks=true]{hyperref}

\usepackage{amssymb}
\usepackage{amsfonts}
\usepackage{t1enc}
\usepackage{pifont}
\usepackage{color}
\usepackage{setspace}

\newtheorem{theo}{Theorem}[section]
\newtheorem{defi}[theo]{Definition}

\newtheorem{corol}[theo]{Corollary}
\newtheorem{prop}[theo]{Proposition}
\newtheorem{lemm}[theo]{Lemma}

\newenvironment{proof}{\noindent{\bf Proof.}\,}{\,$\Box$\par\smallskip}

\newenvironment{rmk}{\par\smallskip\noindent{\it Remark.}\,}{\par\smallskip}
\newcounter{Assumption}
\newcounter{Hypothesis}
\def\blfootnote{\xdef\@thefnmark{}\@footnotetext}
\long\def\symbolfootnote[#1]#2{\begingroup%
\def\thefootnote{\fnsymbol{footnote}}\footnote[#1]{#2}\endgroup}

\newcommand{\Q}{\ensuremath{\mathbb{Q}}}

\newcommand{\R}{\ensuremath{\mathbb{R}}}

\renewcommand{\epsilon }{\varepsilon }
\newcommand{\F}{{\cal F}}

\newcommand{\reels}{{\mathbb{R}}}

\newcommand{\e}{{\rm I\!E}}
\newcommand{\p}{{\rm I\!P}}
\newcommand{\fini}{{\rm I\!F}}

\newcommand{\1}{{\mbox{$1$}\rm \!l}}
\newcommand{\cl}{\mbox{\large $\mathcal{L}$}}

\newcommand{\cs}{\mbox{\large $\mathcal{S}$}}

\newcommand{\f}{\mathcal{F}}
\newcommand{\g}{\mathcal{G}}

\newcommand{\dis}{\displaystyle}

\newcommand{\w}{\widetilde}
\newcommand{\ol}{\overline}

\begin{document}
\begin{frontmatter}

\title{Hedging of Defaultable Contingent Claims using BSDE with uncertain time horizon}

\author{Christophette \textsc{Blanchet-Scalliet}}
\address{Universit\'e de Lyon, CNRS, UMR 5208, Institut Camille Jordan, Ecole Centrale de Lyon, Universit\'e Lyon 1, INSA de Lyon, 36 avenue Guy de Collongue, 69134 Ecully - FRANCE}
\author{Anne \textsc{Eyraud-Loisel}}
\address{Université de Lyon, Laboratoire SAF, ISFA, Universit\'e Lyon $1$, 50 avenue Tony Garnier,69007 Lyon - FRANCE - corresponding author: anne.eyraud@univ-lyon1.fr}
\author{Manuela \textsc{Royer-Carenzi}}
\address{LATP, UMR CNRS 6632 FR 3098 IFR 48 , Evolution Biologique et Modélisation, Université de Provence , Case 19, Pl. V. Hugo , 13331 Marseille Cedex 03 - FRANCE}

\date{\today}
\maketitle

\begin{abstract}
{\small
This article focuses on the mathematical problem of existence and uniqueness of BSDE with a random terminal time which is a general random variable but not a stopping time, as it has been usually the case in the previous literature of BSDE with random terminal time. The main motivation of this work is a financial or actuarial problem of hedging of defaultable contingent claims or life insurance contracts, for which the terminal time is a default time or a death time, which are not stopping times. We have to use progressive enlargement of the Brownian filtration, and to solve the obtained BSDE under this enlarged filtration. This work gives a solution to the mathematical problem and proves the existence and uniqueness of solutions of such BSDE under certain general conditions. This approach is applied to the financial problem of hedging of defaultable contingent claims, and an expression of the hedging strategy is given for a defaultable contingent claim.
}
\end{abstract}

\end{frontmatter}






\section*{Introduction}

In the present work, we study backward stochastic differential equations with uncertain time horizon: the terminal time of the problem is a random variable $\tau$, which is not a stopping time, as usually stated in the previous literature. In our study, $\tau$ is a general random variable.  Hedging problems for defaultable contingent claims fit into this framework, as the terminal time is a default time, which is not a stopping time.

BSDEs were first introduced by E. Pardoux and S. Peng in 1990 \cite{PP90}. These equations naturally appear when describing hedging problems of financial instruments (see \cite{KPQ97} for example). BSDEs with random terminal horizon were introduced by S. Peng (1991) \cite{Pen91} in the Brownian setting, and by E. Pardoux (1995) \cite{Par95} for BSDEs with Brownian setting and Poisson jumps, and were developed by R. Darling and E. Pardoux (1997) \cite{DP97}, P. Briand and Y. Hu (1998) \cite{BH98}, E. Pardoux (1999) \cite{Par99}, M. Royer (2004) \cite{moi04} among others.
The framework of all these studies extensively uses the hidden hypothesis that the processes driven by the BSDE are adapted to the natural Brownian filtration (or Poisson-Brownian in cases with jumps).
As the terminal horizon of our problem is not a stopping time, the filtration that appears to be convenient to work with is not the Brownian filtration ${\cal F}_t$, but the smallest filtration that contains ${\cal F}_t$ and that makes $\tau$ a stopping time. This method is known as progressive enlargement of filtration. It has been introduced in T. Jeulin (1980) \cite{Jeu80b}, T. Jeulin and M. Yor (1978,1985) \cite{JeuYor78b,JeuYor85}, and further developed in J. Azema, T. Jeulin, F. Knight and M. Yor (1992) \cite{AzeJeuYor92}. This framework has been extensively used in default risk models, as the default time is not a stopping time. Works on default risk models have been developed by C. Blanchet-Scalliet and M. Jeanblanc (2004) \cite{blanchet04}, T. Bielecki, M. Jeanblanc and M. Rutkowski (2004) \cite{BiJbRu04b}, M. Jeanblanc and Y. Le Cam (2007) \cite{JeanLeCam07,JeanLeCam08}. Existence of solutions of BSDE under enlarged filtration has been studied by A. Eyraud-Loisel (2005) \cite{Anne-these,Anne2003} for deterministic horizon, and by A. Eyraud-Loisel and M. Royer-Carenzi (2006) \cite{AnnemoiPP1} for random terminal stopping time, under an initially enlarged filtration, as used for asymmetrical information and insider trading modeling.

In a first part, we introduce the model. In a second part, the problem of existence and uniqueness of the BSDE under enlarged filtration ${\cal G}$ is solved. Last section is devoted to an application of previous results to hedging against a defaultable contingent claim. We give an explicit hedging strategy in the defaultable world, under traditional hypothesis ${\bf (H)}$.

\section{Model}
%
%
Let $(\Omega, \fini, \p)$ be a complete probability space and $(W_t)_{0\leq t\leq T}$ be a $m$-dimensional
Brownian motion defined on this space with $W_0=0$. $\f=(\f_t)_{0\leq t\leq T}$ denotes the completed $\sigma$-algebra generated by $W$.\\
We consider a financial market with a riskless asset $S_t^0$ and $m$ risky financial assets $S_t^i$. Prices are supposed to evolve according to the following dynamics :
\begin{equation}
\label{dynamiqueS0}
dS_t^0=r_tS_t^0 \, dt, \quad 0 \leq t \leq T,
\end{equation}
\begin{equation}
\label{dynamiqueS}
dS_t^i=\mu_t^iS_t^i \, dt+S_t^i(\sigma_t^i , dW_t), \quad 0 \leq t \leq T, \forall 1\le i \le m,
\end{equation}
where $r_t \ge 0$ is the risk-free rate, bounded and deterministic, $\mu^i_t$ is the $i$th component of a predictable and vector-valued map $\mu: \Omega \times [0,T] \to \reels^m$ and $\sigma^i_t$ is the $i$th row of a predictable and matrix-valued map $\sigma: \Omega\times[0,T] \to \reels^{m\times m}$.\\
In order to exclude arbitrage opportunities in the financial market we assume that the number of assets is the same as the Brownian dimension. For technical reasons we also suppose that
\begin{enumerate}
\item[(M1)] $\mu$ is bounded and deterministic,
\item[(M2)] $\sigma$ is bounded, in the sense that there exist constants $0 < \epsilon < K$ such that
$\epsilon I_m \le \sigma_t \sigma^*_t \le K I_m$ for all $t\in[0,T]$,
\item[(M3)] $\sigma$ is invertible, and $\sigma^{-1}$ is also bounded.
\end{enumerate}
where $\sigma^*_t$ is the transpose of $\sigma_t$, and $I_m$ is the $m$-dimensional unit matrix.\\
In other words, we require usual conditions to have an arbitrage-free market (\cite{Kar96}), called the the default-free, and even complete market. These conditions ensure the existence of a unique equivalent martingale measure (e.m.m.), denoted by $\tilde{\p}$.\\

Suppose that a financial agent has a positive $\f_0$-measurable initial wealth $X_0$ at time $t=0$. Her wealth at time $t$ is denoted by $X_t$. We consider a hedging problem, represented by a pay-off $\xi$, to be reached under a random terminal condition, which is not a stopping time. It is the case for defaultable contingent claims, where the terminal time is a default time.
For example, an agent sells an option with maturity $T$, based on a defaultable asset. This type of contract (defaultable contingent claim) generally leads to two possible payoffs: the seller commits itself to give the payoff of a regular option, if default did not occur at time $T$, which will be represented by a $\F_T$-measurable random variable $V$ (for instance, $V=(S_T-K)_+$ for a european call option, but in general, $V$ may depend on the paths of asset prices until time $T$); if default occurs before time $T$, the seller has to pay at default time a compensation $C_\tau$, defined as the value at default time $\tau$ of an $\f_t$-predictable nonnegative semi-martingale $C_t$.\\
Then the final payoff at time $\tau \wedge T$ has the general form~:
%
%
\begin{eqnarray*}
\xi & = & V \, \1_{\tau >T} + C_\tau \, \1_{\tau \le T},
\end{eqnarray*}

Default times are random variables that do not depend entirely on the paths of some financial risky assets. They may have a financial component, but have an exogenous part, which makes them not adapted to the natural filtration generated by the observations of prices.\\
Nevertheless, they are observable : at any time, the common agent can observe if default $\tau$ has occurred or not. The information of an agent is therefore not the filtration generated by the price processes $(\f_t)_{0\leq t\leq T}$, but is defined by ${\cal G}= ({\cal G}_t)_{t\in[0,T]}$, where
\begin{equation}
{\cal G}_t =  {\cal F}_t\vee \sigma(\1_{\tau\le t}),
\end{equation}
which is the completion of the smallest filtration that contains filtration $(\f_t)_{0\leq t\leq T}$ and that makes $\tau$ a stopping time. So the previous payoff belongs to the following space :
$
 \xi \in {\cal G}_{T\wedge \tau}.
$
The problem is to find a hedging admissible strategy, i.e. a strategy that leads to the terminal wealth $X_{T\wedge \tau}=\xi$.\\

Under ${\cal G}$, the default-free market is not complete any more. The martingale representation property has to be established under this new filtration. For short, to be able to hedge against the random time, another asset will be needed, in order to fill up the martingale representation property.

In financial defaultable markets, the payment of a contingent claim depends on the default occurrence before maturity. Therefore another tradable asset (or at least attainable) is often considered : the defaultable zero-coupon bond with maturity $T$, whose value at time $t$ is $\rho_t=\rho(t,T)$. This asset will give its owner the face-value $1$ if default did not occur before $T$, and nothing otherwise.\\

If this asset is tradable on the market, an admissible hedging strategy will be a self-financing strategy based on the non risky asset, the risky asset, and the defaultable zero-coupon.

\section{Solution of the BSDE under ${\cal G}$}
\label{BSDEsec}

To avoid arbitrage opportunities, we work in a mathematical set up where $(\f, \p)$ semi-martingales remain $(\g, \p)$ semi-martingales. This property does not hold at any time. In context of credit risk, the good hypothesis consists in supposing that $\tau$ is an initial time; it is called Density Hypothesis, detailed by M. Jeanblanc and Y. Le Cam in \cite{JeanLeCam08} and also by N. El Karoui \textit{et al.} in \cite{ElkJeanJiao}.\\

\textbf{Density Hypothesis} : We assume that there exists an $\f_t \times \mathcal{B} (\R^+)$-measurable function $\alpha_t \, : \, (\omega,\theta)\rightarrow \alpha_t(\omega, \theta)$ which satisfies
\[
\p(\tau\in d\theta |{\cal F}_t): =\alpha_t(\theta) \, d\theta, \, \p-a.s.
\]
\\
\begin{rmk}
For any $\theta$, the process $\left(\alpha_t(\theta)\right)_{ t\leq 0}$ is an $(\f,\p)$ non-negative martingale.

\end{rmk}
We introduce the following conditional probability
\begin{eqnarray}
\label{defF}
F_t=\e_{\p} (\1_{\tau\le t}|{\cal F}_t) & = & {\p}(\tau \leq t|{\cal F}_t).
\end{eqnarray}
We will always consider the right-continuous version of this $(\f, \p)$-submartingale, and we will also assume that $F_t<1$ a.s. $\forall t\in [0,T]$. Define the $\f$-predictable, right-continuous nondecreasing process $(\hat{F}_t)_{t \geq 0}$ such that the process $F-\hat{F}$ is a $(\f, \p)$-martingale, denoted by $(M_t^F)_{t\ge 0}$. We denote by $(\psi)_{t\ge 0}$ the process such that $dM_t^F=\psi_t \, dW_t$.\\[5pt]

Under the Density Hypothesis, it is well known that
\[
F_t=\int_0^{t}\alpha_t(s) \, ds
\]
and that the process
\[
M^{}_t  = H_t-\int_0^{t\wedge \tau}(1-H_s) \, \frac{\alpha_s(s)}{1-F_s} \, ds
\]
is a $(\g, \p)$-martingale, where process $(H_t)_{t \geq 0}$ is the defaultable process with $H_t=\1_{\tau\le t}$, and process $(\lambda_t)_{t \geq 0}$ is defined by $\lambda_t=\frac{\alpha_t(t)}{1-F_t}$ (see \cite{BiRu02} and \cite{JeanLeCam08}).


\subsection{Representation theorem}

In such a context any $(\f, \p)$-martingale $X$ is a $(\g, \p)$ semi-martingale and the process $\bar{X}$ defined by
\begin{equation}
\label{barre}
\bar{X}_{t} \ = \ X_{t} \, - \, \int_0^{t\wedge \tau}{\frac{d\left\langle X,F\right\rangle_s}{1-F_{s^-}}} \, - \, \int_{t\wedge \tau}^t{\frac{d\left\langle X,\alpha(u)\right\rangle_s}{\alpha_{s^-}(u)}\left|_{u=\tau}\right.},  \quad 0 \leq t \leq T\,
\end{equation}

is a $(\g, \p)$-martingale (see M. Jeanblanc and Y. Le Cam in \cite{JeanLeCam09}).

$(W_t)_{t \geq 0}$ is a Brownian motion in probability space $(\Omega, \f, \p)$, and we denote by $\bar{W}$ the associated Brownian motion under $(\Omega, \g, \p)$, defined by Equation (\ref{barre}).\\

For any $\gamma \in \reels$, let us define ${\cal B}_{\gamma}^2={\cal S}_{\gamma}^2\times
\cl^2_{\gamma}(\bar{W}, \p)\times \cl^2_{\gamma}( M, \p)$ where we denote by :
\begin{itemize}
\item  ${\cal S}^2_{\gamma}$ the set of $1$-dimensional ${\cal G}$-adapted c\`adl\`ag
processes $(Y_t)_{0\le t\le T}$\\
such that $\dis ||Y||_{{\cal S}_{\gamma}^2}^2= { \e_{\p} \Big( \sup_{0\le t \le T} { e^{\gamma \, (t \wedge \tau)} \, Y_{t\wedge \tau}^2} \Big)} <\infty $,
\item $\cl^2_{\gamma}(\bar{W}, \p)$ the set of all $m$-dimensional
${\cal G}$-predictable processes $(Z_t)_{0\le t\le T}$ such that
$||Z||_{\cl^2_{\gamma}(\bar{W}, \p)}^2=\e_{\p} \Big( \int_0^{T\wedge \tau}{ e^{\gamma \, s} \, \|Z_s\|^2\, ds \Big) }<\infty$,
\item $\cl^2_{\gamma}(M, \p)$ the set of all $1$-dimensional
${\cal G}$-predictable processes $(U_t)_{0\le
t\le T}$ such that $||U||_{\cl^2_{\gamma}(M, \p)}^2= \e_{\p} \Big( \int_0^{T\wedge \tau} e^{\gamma \, s} \, |U_s|^2 \, \lambda_s \, ds \Big) <\infty$.
\end{itemize}
Let recall a representation theorem established by Jeanblanc and Le Cam under "density hypothesis"(see theorem 2.1 \cite{JeanLeCam08})

\begin{theo}
\label{Yorrepre}
For every $(\g,\p)$ martingale $\bar{X}$, there exist two $\g$-predictable process $\beta$ and $\gamma$ such that
\[
d\bar{X}_t \ = \ \gamma_t \, d\bar{W}_t \, + \, \beta_t \, dM_t
\]
\end{theo}

\begin{rmk}
If $\bar{X}$ is square integrable martingale, then the process $\gamma$ (respectively $\beta$) belongs to $\cl^2_{\gamma}(\bar{W}, \p)$ (resp. $\cl^2_{\gamma}(M, \p)$).
\end{rmk}


\subsection{Existence theorem}
\label{sectionThH3cont}


Fix $T>0$ and $\xi \in \cl^2( \g_{T \wedge \tau})$.

\noindent The BSDE to be solved is the following :
\begin{equation}
\label{EDSRale}
Y_{t \wedge \tau} =\xi +\int_{t \wedge \tau} ^{T \wedge \tau} {f(s,Y_s,Z_s,U_s)\,
ds}-\int_{t \wedge \tau} ^{T \wedge \tau} {Z_s \, d\bar{W}_s}-\int_{t \wedge \tau} ^{T \wedge \tau} {U_s\,dM^{}_s},\ \ 0\le t\le T.
\end{equation}
The aim of this section is to prove an existence and uniqueness result for this BSDE stopped at $\g$-stopping time $T \wedge\tau$.
In the previous financial interpretation, this unique $\g$-adapted solution $(Y,Z,U)$, stopped at time $\tau$, will represent the unique portfolio that hedges the defaultable contingent claim.

\textbf{Hypotheses on $f$ and $\lambda$} :
\begin{itemize}
\item $\lambda$ is a non-negative function, bounded by a constant $K_1$ ;
\item f is a Lipchitz function such that there exist a constant $K_2$ satisfying
\begin{eqnarray}
\left|f(s,y,z,u)-f(s,y',z',u')\right|\leq K_2 \, (\left|y-y' \right|+\left\|z-z' \right\|)+ \lambda_s\left|u-u' \right| .
	\label{conditionf}
\end{eqnarray}
Let us denote $K= max(K_1, K_2)$.
\end{itemize}

\begin{defi}
\label{defi ale}
~\\
Let us consider $T >0$ and $\xi \in \cl^2 (\Omega,\g_{T \wedge \tau},\p)$. A $(\Omega,\g,\p)$-solution (or a solution on $(\Omega,\g,\p)$) to equation (\ref{EDSRale}) is a triple of  $\reels \times \reels ^{m} \times \reels $-valued $\big( Y_t,Z_t, U_t \big)_{t \geq 0}$ processes such that
\begin{enumerate}
\item $Y$ is a ${\cal G}$-adapted c\`adl\`ag
process and $(Z,U) \in \cl^2_{0}(\bar{W}, \p)\times \cl^2_0( M, \p) $,
\item On the
set $\{  t \geq T \wedge \tau \}$, we have $ Y_t = \xi, \, Z_t = 0 $ and $U_t = 0$,
\item $\forall \, r \in [0,T] \ and \ \forall t \in [0, r]$, we have\\[2pt]
 $Y_{t\wedge \tau} \, = \, Y_{r\wedge \tau} +\int_{t\wedge \tau}^{r\wedge \tau}{f(s,Y_s,Z_s, U_s) \, ds} - \int_{t\wedge \tau}^{r\wedge \tau}{Z_s \, d\bar{W}_s} -\int_{t \wedge \tau} ^{r \wedge \tau} {U_s \, dM_s}$.
\end{enumerate}
\end{defi}

\begin{lemm}
\label{batterie}
~\\
Let $\xi\in \cl^2(\Omega,{\cal G}_{T\wedge \tau},\p)$. If $(Y_t,Z_t, U_t)_{0\le
t\le T}$ is a $(\Omega, {\cal G},\p)$-solution of
the BDSE (\ref{EDSRale}) as defined in the Definition \ref{defi ale}, with $f$ satisfying hypothesis (\ref{conditionf}) and
\[
\e_{}\left(\int_0 ^{T \wedge \tau}\left|f(s,0,0,0)\right|^2ds\right) < +\infty,
\]
then \vspace{-.2cm}
\begin{displaymath}
\e_{}\left( \sup_{0\le t\le T} \ Y_{t\wedge \tau}^2\right) < +\infty.
\end{displaymath}
\end{lemm}

\begin{proof}
~\\
The proof is given in Appendix.
\end{proof}

We can now state the following theorem :
\begin{theo}
\label{Tfixe}
~\\
Let $\xi \in \cl^2(\Omega,{\cal G}_{T\wedge \tau},\p)$ and $f
:\Omega\times[0,T]\times{\reels}\times{\reels}^{m}\times{\reels}\longrightarrow {\reels}$ be $\g$-measurable.\\
If \ $\e \left( \int_0^T|f(s,0,0,0)|^2 \, ds \right) <\infty$ and if $f$ satisfies condition (\ref{conditionf}), there exists a unique $\g$-adapted triple $(Y,Z,U)\in {\cal
B}^2_{0}$ solution of the BSDE:
$$
Y_{t \wedge \tau} =\xi +\int_{t \wedge \tau} ^{T \wedge \tau} {f(s,Y_s,Z_s,U_s)\,
ds}-\int_{t \wedge \tau} ^{T \wedge \tau} {Z_s \, d\bar{W}_s}-\int_{t \wedge \tau} ^{T \wedge \tau} {U_s\, dM_s},\ \ 0\le t\le T.
$$
\end{theo}
%
\begin{proof}
~\\
We can adopt the usual contraction method using representation Theorem \ref{Yorrepre}.\\
Let $\gamma \in \reels$. Recall that ${\cal B}_{\gamma}^2={\cal S}^2_{\gamma} \times \cl^2_{\gamma} (\bar{W}, \p) \times \cl^2_{\gamma}( M, \p)$. We define a function
$\Phi: {\cal B}^2_{0} \rightarrow {\cal B}^2_{0}$ such that $(Y,Z,U)\in {\cal B}^2_{0}$ is a solution of our BSDE if it is a fixed point of $\Phi$.\\
Let $(y,z,u) \in {\cal B}^2_{0}$. Define $(Y,Z,U)=\Phi (y,z,u)$ with :\\
$$Y_t = \e \Big( \xi + \int_{t \wedge \tau}^{T \wedge \tau} f(s,y_s,z_s,u_s) \, ds \ \Big|\ \g_t \Big) \mbox{ , }0\leq t\leq T \mbox{ , }$$
and processes $(Z_t)_{0\leq t\leq T} \in \cl^2_{0}(\bar{W}, \p)$ and $(U_t)_{0\leq t\leq T} \in \cl^2_{0}( M, \p)$ obtained by using martingale representation Theorem
\ref{Yorrepre} applied to the square integrable ($\g, \p$)-martingale $(N_t)_{0 \leq t \leq T}$ where $N_{t} = \e \Big(  \xi+ \int_0^{T\wedge \tau}{f(s,y_s,z_s,u_s) \, ds} \ \Big|\ \g_t \Big)$.\\ Hence
$$
N_{t \wedge \tau} = N_{T \wedge \tau} - \int_{t \wedge \tau} ^{T \wedge \tau} {Z_s\, d\bar{W}_s} - \int_{t \wedge \tau} ^{T \wedge \tau}{U_s\, dM_s},
$$
$$
Y_{t \wedge \tau} \, + \, \int_0^{t \wedge \tau}  {f(s,y_s,z_s,u_s) \, ds} \ = \ \xi \, + \, \int_0^{T\wedge \tau}{f(s,y_s,z_s,u_s) \, ds}\\[5pt]
$$
$$
- \, \int_{t \wedge \tau} ^{T \wedge \tau} {Z_s\, d\bar{W}_s} \, - \, \int_{t \wedge \tau} ^{T \wedge \tau}{U_s\, dM_s}.
$$
Consequently
$$
Y_{t \wedge \tau} \ = \ \xi+ \int_{t \wedge \tau} ^{T\wedge \tau}{f(s,y_s,z_s,u_s) \, ds} - \int_{t \wedge \tau} ^{T \wedge \tau} {Z_s\, d\bar{W}_s} - \int_{t \wedge \tau} ^{T \wedge \tau}{U_s\, dM_s}.
$$
This means that $(Y,Z,U)$ is a $(\,\Omega,\, \g \, ,\, \p\,)$-solution to Equation (\ref{EDSRale}) with particular generator $s \mapsto g(s) = f(s,y_s,z_s,u_s)$, which implies thanks to Lemma \ref{batterie} that the triple $(Y,Z,U)$ belongs to the convenient space $\mathcal{B}^2_{0}$ and consequently map $\Phi$ is well defined.

\noindent Next, for $(y^1, z^1, u^1)$ and $(y^2, z^2, u^2)$ in $ {\cal B}_{0}^2$, we define $(Y^1,Z^1,U^1)=\Phi (y^1, z^1, u^1)$ and $(Y^2,Z^2, U^2)=\Phi (y^2, z^2, u^2)$. Let $(\hat{y},\hat{z},\hat{u}) = ({y^1} - {y^2}, {z^1} - {z^2}, {u^1} - {u^2})$ and $(\hat{Y},\hat{Z},\hat{U}) = ({Y^1} - {Y^2}, {Z^1} - {Z^2}, {U^1} - {U^2})$.\\[5pt]
Then
\begin{eqnarray*}
\hat{Y}_{t \wedge \tau} & = & \int_{t \wedge \tau} ^{T\wedge \tau}\left( f(s,y^1_s,z^1_s,u^1_s) \,- f(s,y^2_s,z^2_s,u^2_s) \right) \, ds\\[5pt]
 & & - \, \int_{t \wedge \tau} ^{T \wedge \tau} \hat{Z}_s \, d\bar{W}_s \, - \, \int_{t \wedge \tau} ^{T \wedge \tau} \hat{U}_s \, dM_s.
\end{eqnarray*}
Let us apply It\^o's formula to process $\left( e^{\gamma \, t} \, Y_t^2 \right)_{0 \leq t \leq T}$. Taking $\gamma=4K^2+2K+1$, it gives for any $t$ in $[0,T]$ :
\begin{eqnarray*}
\lefteqn{\e \left( \int_{t \wedge \tau}^{T\wedge \tau}{e^{\gamma s} \, ( \, \hat{Y}_s^2+ \| \hat{Z}_s\|^2 \, ) \, ds} \, + \, \int_{t \wedge \tau}^{T\wedge \tau} e^{\gamma s} \, \hat{U}_s^2 \, \lambda_s \, ds \, \right)}\\[10pt]
&\leq & \frac{1}{2} \, \e \left( \int_0^{T\wedge \tau}{e^{\gamma s} \, ( \hat{y}_s^2+\|\hat{z}_s\|^2 ) \, ds} + \int_0^{T\wedge \tau} e^{\gamma s} \, \hat{u}_s^2 \, \lambda_s \, ds \right).
\end{eqnarray*}
And finally, with $t=0$,
\begin{eqnarray*}
\lefteqn{\e \left( \int_0^{T\wedge \tau}{e^{\gamma s} \, ( \, \hat{Y}_s^2+ \| \hat{Z}_s\|^2 \, ) \, ds} \, + \, \int_0^{T\wedge \tau} e^{\gamma s} \, \hat{U}_s^2 \, \lambda_s \, ds \, \right)}\\[10pt]
&\leq & \frac{1}{2} \, \e \left( \int_0^{T\wedge \tau}{e^{\gamma s} \, ( \hat{y}_s^2+\|\hat{z}_s\|^2 ) \, ds} + \int_0^{T\wedge \tau} e^{\gamma s} \, \hat{u}_s^2 \, \lambda_s \, ds \right).
\end{eqnarray*}
Then $\Phi$ is a strict contraction on ${\cal B}^2_{0}$ with norm\\
$$|||(Y,Z,U)|||_\gamma=  \e\left( \int_0^{T\wedge \tau}{e^{\gamma s} \, ( {Y}_s^2+\| {Z}_s\|^2 ) \, ds} + \int_0^{T\wedge \tau} e^{\gamma s} \, U_s^2 \, \lambda_s \, ds \right) ^{\frac{1}{2}}$$.\\[2pt]
We finally deduce that $\Phi$ has a unique fixed point and conclude that the BSDE has a unique solution.
\end{proof}
%


\section{Hedging strategy in the defaultable world with BSDE}


\subsection{Defaultable zero-coupon}


After giving in Section \ref{BSDEsec} the results in a framework of initial times, we restrict hereafter to consider the particular case where

\[\alpha_t(u)= \alpha_u(u), \quad \forall u \leq t\]

This case is equivalent to the hypothesis called immersion property or Hypothesis \textbf{(H)}.

\paragraph{\textbf{Hypothesis}{ \bf(H)}} {\it Any square integrable (${\cal F}, \, \p$)-martingale is a square integrable $({\cal G}, \, \p)$-martingale}.\\

\noindent Under this hypothesis, the process $F$ is continuous and Brownian motion $W$ is still a Brownian motion in the enlarged filtration. The results obtained in the previous section are still satisfied, with $W$ instead of $\bar{W}$.
As explained in the introduction, we denote by $\tilde{\p}$ the unique e.m.m equivalent to $\p$ on ${\cal F}$.
According to section 3.3 in \cite{blanchet04}, when ${\bf(H)}$ holds on the historical probability, as soon as the $\f$-market is complete, the defaultable market is still arbitrage free. $\textbf{(H)}$ holds under any $ \g$-equivalent martingale measure ${\p}^{\psi}$ such that ${\p}^{\psi}_{| \mathcal{G}_t} \ = \ K^{\psi}_t \, \p_{| \mathcal{G}_t}$ with
\begin{displaymath}
dK^{\psi}_t=K^{\psi}_{t^-}(-\theta_t \, dW_t+\psi_t \, dM_t) \, , \quad 0 \leq t \leq T,
\end{displaymath}
where $\theta=\sigma^{-1}(\mu-r)$ denotes the risk premium and $\psi>-1$.

The equation satisfied by $K^{\psi}$ is obtained using a representation theorem for all $(\g,\p)$ square-integrable martingales established by S. Kusuoka \cite{Kusu99} under hypothesis \textbf{(H)}.\\
Let ${\p}^{\psi}$ be such a $ \g$-equivalent martingale measure. We have ${\p}^{\psi}_{| \f}=\p^0_{| \f}=\tilde{\p}_{| \f}$. $W^{0}$ denotes the Brownian motion obtained using Girsanov's transformation (since the coefficient in the Radon-Nikod\'ym density associated to the Brownian motion is always  $\theta$). We also introduce processes $F^{\psi}$ and $M^{\psi}$ constructed in the same way as $F$ and $M$ but associated to the probability $\p^{\psi}$ instead of $\p$. Note that process $F^{\psi}$ is continuous because $\tau$ is still an initial time with immersion property under $\p^{\psi}$ (see M. Jeanblanc and Y. Le Cam in \cite{JeanLeCam08}). Then using Girsanov's transformation, the $(\g,\p^{\psi})$-martingale $M^{\psi}$ satisfies $dM^{\psi}_t \ = \ dM_t \, - \, (1-H_t) \, (1+\psi_t) \, \lambda_t \, dt$.\\[5pt]
Let  $( \w{\rho_t})_{0 \leq t \leq T}$ be the discounted price of the defaultable zero-coupon bond and $R_t$ the discount factor :
\begin{eqnarray*}
R_t & = & \exp\left(-\int_0^t r_s \, ds \right), \quad 0 \leq t \leq T.
\end{eqnarray*}
We obtain from Proposition $2$ in \cite{blanchet04} the following result :
\begin{lemm}
\begin{displaymath}
d\w{\rho_t} \ = \ \frac{\1_{\tau > t}}{1- F^{\psi}_{t^-}} \, \phi^m_t \, dW^{0}_t \, - \, \w{\rho}_{t^-} \, dM^{\psi}_t \,  , \quad 0 \leq t \leq T,
\end{displaymath}
\end{lemm}
\paragraph{Proof} $(\phi^m_t)_{t \geq 0}$ comes from the representation of $({\cal F},
{\p^{0}})$-martingale $(m_t)_{t \geq 0} \ = $\\
$\Big( \e_{\p^{0}}(R_T \, \1_{\tau > T} \, | \, {\cal F}_t) \Big)_{t \geq 0}$ with respect to
$({\cal F}, {\p^{0}})$-Brownian motion ${W^{0}}$.\\
As $\forall t\in ]0,T\wedge \tau]\,\  \w{\rho}_{t^-} \neq 0$, we can set $\dis c_t \ = \ \frac{\1_{\tau > t}}{1-F^{\psi}_{t^-}} \, \frac{\phi^m_t}{\w{\rho}_{t^-}} $.

Using Girsanov transformation, we obtain finally the dynamics of the defaultable zero-coupon under historical probability :
\begin{prop}
\begin{equation}
\label{dynamiquezerocoupon}
d\rho_t \ = \  \rho_{t^-} \, (a_t \, dt \, + \, c_t \, dW_t \, - \, dM_t),
\end{equation}
where :
\begin{equation}
\label{nonarbitrage}
a_t \ = \ r_t \, +\, \theta_t \, c_t \, + \, (1-H_{t^-}) \, \psi_t \, \lambda_t .
\end{equation}
\end{prop}


\subsection{Wealth's dynamic}


\subsubsection{BSDE formulation}


Let $Y_t$ be the wealth at time $t$ of the agent. Suppose that she has $\alpha_t$ parts of the risky asset, $\delta_t$ parts of the riskless asset, and $\beta_t$ parts of the defaultable zero-coupon bond. At any time $t$, we have :
\begin{equation}
\label{repartition}
Y_t \ = \ \alpha_t \, S_t+\beta_t \, \rho_{t^-}+\delta_t \, S_t^0.
\end{equation}
where $\alpha_t, \beta_t$ and $\delta_t$ are predictable.\\
The self-financing hypothesis can be written as :
\begin{eqnarray*}
dY_t &=& \alpha_t \, dS_t+\beta_t \, d\rho_t+\delta_t \, dS_t^0,
\end{eqnarray*}
which can be developed, for any $t$ in $[0, T \wedge \tau]$, using  (\ref{repartition}) and the dynamics of the three assets (\ref{dynamiqueS}), (\ref{dynamiquezerocoupon}) and (\ref{dynamiqueS0}). This yields to
\begin{eqnarray*}
dY_t & = & \left(\alpha_t \, \mu_t \, S_t+r_t \, Y_t-\alpha_t \, r_t \, S_t-\beta_t \, r_t \, \rho_{t^-} + \beta_t \, a_t \, \rho_{t^-} \right) \, dt \ \\[3pt]
& & + \left( \alpha_t \, \sigma_t \, S_t + \beta_t \, c_t \, \rho_{t^-} \right) \, dW_t - \beta_t \, \rho_{t^{-}} \, dM_t.
\end{eqnarray*}
Then, denoting by $Z_t \ = \ \alpha_t \, \sigma_t \, S_t + \beta_t \, c_t \, \rho_{t^-}$ and $U_t \ = \ - \, \beta_t \, \rho_{t^{-}} $, we obtain a BSDE satisfied by the wealth process $Y_t$ :
\begin{equation}
\label{edsrrichesse}
\left\{
\begin{array}{lll}
dY_t & = & -f(t,Y_t,Z_t,U_t)\, dt +Z_t \, dW_t+U_t \, dM_t, \ 0 \leq t \leq T \wedge \tau\\
Y_{T \wedge \tau} & = & \xi
\end{array}
\right.
\end{equation}
with $f(t,y,z,u) \ = \ -r_t\, y \, - \,  \theta_t \, z
 + \left( \, a_t \, - \, r_t \, - \, \theta_t \, c_t \, \right) \, u $.\\
%
Using (\ref{nonarbitrage}), we obtain
\begin{eqnarray}
\label{fappfin}
f(t,y,z,u) & = & -r_t\, y - \theta_t \, z + \, (1-H_{t^-}) \, \psi_t \, \lambda_t\, u.
\end{eqnarray}
This provides a BSDE with ${\cal G}_t$-adapted coefficients.
As $\f$-Brownian motion $W$ is still a Brownian motion under the new filtration $\g$, the previous
stochastic differential equation has a sense.

%


\subsubsection{Application of Theorem \ref{Tfixe}}

As condition (\ref{conditionf}) holds true, as $r, \theta$ and $\lambda$ are bounded, and as $f(s,0,0,0)=0$, the integrability condition on $f$ under $\p$ is also satisfied, Theorem \ref{Tfixe} guarantees existence and uniqueness of the solution of the previous BSDE.

\begin{prop}
There exists a unique solution of BSDE (\ref{edsrrichesse}) with driver (\ref{fappfin}), for all $\xi \in \cl^2({\cal G}_{T\wedge\tau})$.
\end{prop}


\subsubsection{Explicit solution for the hedging strategy}
%
%
When $\xi = V \, \1_{\tau> T} + C_{\tau} \, \1_{\tau \leq T}$ represents a defaultable contingent claim, we give an explicit solution for the hedging strategy, given by the solution of (\ref{edsrrichesse}) with driver (\ref{fappfin}).

%

\begin{theo}
\label{Lyon}
~\\
Let $V \in \cl^2(\f_T)$ and $C$ be a square integrable $\f$-predictable process.  $$\xi = V \, \1_{\tau> T} + C_{\tau} \, \1_{\tau \leq T}$$
Let  $f:\Omega\times[0,T]\times \reels \times \reels^{m}\times \reels \longrightarrow \reels$ be the $\g$-measurable generator defined by $$f(t,y,z,u) = -r_t\, y - \theta_t \, z +\, (1-H_{t^-}) \, \psi_t \, \lambda_t\, u,$$ satisfying condition (\ref{conditionf}).\\
Then, under hypothesis {\bf (H)}, there exists a unique $\g$-adapted triple $(Y,Z,U)\in {\cal B}^2_0$ solution of the BSDE :
\begin{equation}
\label{EDSRappFin}
Y_{t \wedge \tau} =\xi +\int_{t \wedge \tau} ^{T \wedge \tau} {f(s,Y_s,Z_s,U_s)\,
ds}-\int_{t \wedge \tau} ^{T \wedge \tau} {Z_s\, dW_s}-\int_{t \wedge \tau} ^{T \wedge \tau} {U_s\, dM_s},\ \ 0\le t\le T.
\end{equation}
Moreover $$Z_t= \frac{a_t^C+ a_t^{V}}{R_t(1-F^{\psi}_t)},$$ and $$U_t=C_t - R_t^{-1} \e_{\p^{\psi}}(R_\tau C_\tau|{\cal G}_{t^-}) - R_t^{-1} \e_{\p^{\psi}}(R_TV \1_{T<\tau}|{\cal G}_{t^-}),$$
where $(a_t^C)_{t \geq 0}$ comes from the representation of $({\cal F}, \p^{\psi})$-martingale\\
$ \bigg( \e_{\p^{\psi}} \left(\int_0^\infty R_sC_s \, dF^{\psi}_s|{\cal F}_t\right) \bigg)_{t \geq 0}$ and $(a_t^{V})_{t \geq 0}$ from $\bigg(  \e_{\p^{\psi}}(R_TV \, \1_{\tau > {T}} \, | \, {\cal F}_t)\bigg)_{t \geq 0}$.
\end{theo}
\begin{proof}
~\\
Let us consider the discounted process $(R_tY_t)_{0\leq t\leq T}$. We have
\begin{eqnarray*}
R_{t \wedge \tau} \, Y_{t \wedge \tau} \ = \  \e_{\p^{\psi}}(  R_{T \wedge \tau} \, \xi | \g_t).
\end{eqnarray*}
We compute separately the conditional expectation of $R_{\tau} C_{ \tau} \, \1_{\tau \leq T}$ and $R_T \, V \, \1_{T < \tau}$.\\
Let $X^C_t= \e_{\p^{\psi}}(R_{\tau}C_\tau\, \1_{\tau \leq T}|{\cal G}_t)$.\\[5pt]
%
From Proposition $3$ in C. Blanchet-Scalliet and M. Jeanblanc \cite{blanchet04}, we have
\begin{equation}\label{representation blanchetC}
X^C_t= X^C_0+\int_0^{t\wedge \tau}\frac{1}{{\p^{\psi}}(\tau > {s} \, | \, {\cal F}_{s})} \,
a_s^C \, dW^0_s+\int_0^{t\wedge \tau} (R_sC_s-X^C_{s^-})\, dM^{\psi}_s,
\end{equation}
where $(a_t^C)_{t \geq 0}$ comes from the representation of the $({\cal F},\p^{\psi})$-martingale\\
$ \bigg( \e_{\p^{\psi}} \left(\int_0^\infty R_s C_s \, dF_s|{\cal F}_t\right) \bigg)_{t \geq 0}$ with respect to
$({\cal F}, \p^{\psi})$-Brownian motion $W^0$.\\[5pt]%
%
For the second term, 
 $X_t^V=\e_{\p^{\psi}}(R_TV \, \1_{T<\tau}|{\cal G}_t)$ is a $({\cal G},\p^{\psi})$-martingale and can be represented as follows :
\begin{equation}\label{representation blanchetV}
X_t^V =X_0^V +\int_0^{t\wedge \tau}\frac{1}{{\p^{\psi}}(\tau > {s} \, | \, {\cal F}_{s})}\,
a_s^{V} \, dW^0_s -\int_0^{t\wedge \tau}X^V_{s^-}\, dM^{\psi}_s,
\end{equation}
where $(a_t^{V})_{t \geq 0}$ comes from the representation of the $({\cal F},\p^{\psi})$-martingale\\
$\bigg(  \e_{\p^{\psi}}(R_T V \, \1_{\tau > {T}} \, | \, {\cal F}_t)\bigg)_{t \geq 0}$ with respect to $({\cal F}, \p^{\psi})$-Brownian motion $W^0$.\\
Summing (\ref{representation blanchetC}) and (\ref{representation blanchetV}), we obtain $R_sZ_s = \frac{a_s^C+ a_s^{V}}{ 1-F^{\psi}_s } $ and $R_sU_s = R_sC_s-X^C_{s^-} - X^V_{s^-} $. \\
Since $X_t^V$ and $X^C_t$ are square integrable, $Z\in \cl^2_{0}(W,\p)$. Using Theorem \ref{Tfixe}, $(Y,Z,U)$ is the unique solution of BSDE (\ref{EDSRappFin}) in $\mathcal{S}^2 \times \mathcal{L}^2_{0} (W, \p) \times \mathcal{L}^2_{0} (M, \p)$.
\end{proof}

\begin{rmk}
By solving BSDEs, we detailed a new approach to find the same results as those stated in C. Blanchet-Scalliet and M. Jeanblanc \cite{blanchet04}, as a special case of the last Theorem.
\end{rmk}
\section{Conclusion}

This article has presented a new BSDE approach to finding hedging strategies in a defaultable world. Results have been obtained for a large panel of hedging payoffs, and under general assumptions. The hedging portfolios have been expressed in term of a solution of a backward stochastic differential equation.

%
%


\nocite{}
\bibliography{ref}
\bibliographystyle{amsplain}

\medskip

\medskip

\appendix{\large{Appendix : Proof of Lemma \ref{batterie}}}

Let $(Y_t,Z_t, U_t)_{0\le t\le T}$ be a solution of (\ref{EDSRale}) :
\begin{displaymath}
Y_{t \wedge \tau} =\xi +\int_{t \wedge \tau} ^{T \wedge \tau} {f(s,Y_s,Z_s,U_s)\,
ds}-\int_{t \wedge \tau} ^{T \wedge \tau} {Z_s \, d\bar{W}_s}-\int_{t \wedge \tau} ^{T \wedge \tau} {U_s\, dM_s},\ \ 0\le t\le T.
\end{displaymath}
Let us consider $\gamma \in \reels $.
Apply Itô's formula to the process $\left(e^{\gamma t} \, Y^2_t\right)_{t \geq0}$ between $t \wedge \tau$ and $T \wedge \tau$.
\begin{eqnarray*}
e^{\gamma (t \wedge \tau)}\, Y^2_{t \wedge \tau} & = & e^{\gamma (T \wedge \tau)} \ \xi ^2 \, - \, \gamma\int_{t \wedge \tau} ^{T \wedge \tau}e^{\gamma s} \, Y_s^2 \, ds \, + \, 2\int_{t \wedge \tau} ^{T \wedge \tau}e^{\gamma s} \, Y_s \, f(s,Y_s,Z_s,U_s) \, ds\\[10pt]
& & \, - \, 2\int_{t \wedge \tau} ^{T \wedge \tau}e^{\gamma s} \, Y_s \, Z_s \, d\bar{W}_s \, - \, 2\int_{t \wedge \tau} ^{T \wedge \tau}e^{\gamma s} \, Y_{s-} \, U_s \,  dM_s\\[10pt]
& &  - \, \int_{t \wedge \tau} ^{T \wedge \tau}e^{\gamma s} \, \left\|Z_s\right\|^2ds \, - \, \sum_ {t \wedge \tau \leq s \leq T \wedge \tau}e^{\gamma s} \, U_s^2 \, \Delta H_s \, . \\
\end{eqnarray*}
Then
\begin{displaymath}
e^{\gamma (t \wedge \tau)} \ Y^2_{t \wedge \tau} \, + \, \gamma\int_{t \wedge \tau} ^{T \wedge \tau}e^{\gamma s} \, Y_s^2 \, ds
\end{displaymath}

\vspace{-0.5cm}
\begin{eqnarray*}
& \leq & e^{\gamma (T \wedge \tau)} \ \xi^2 \, + \, 2\int_{t \wedge \tau} ^{T \wedge \tau}e^{\gamma s} \, Y_s \, f(s,Y_s,Z_s,U_s) \, ds \\[10pt]
 & &  - \, 2\int_{t \wedge \tau} ^{T \wedge \tau}e^{\gamma s} \, Y_s \, Z_s \, d\bar{W}_s \, - \, 2\int_{t \wedge \tau} ^{T \wedge \tau}e^{\gamma s} \, Y_{s^-} \, U_s \, dM^{}_s \, . \\[20pt]
& \leq & e^{\gamma (T \wedge \tau)} \ \xi^2 \, + \, \int_{t \wedge \tau} ^{T \wedge \tau}e^{\gamma s} \, \left|f(s,0,0,0)\right|^2 ds \, + \, (1+3K+K^2) \ \int_{t \wedge \tau} ^{T \wedge \tau}e^{\gamma s} \, Y_s^2 \, ds \\[10pt]
& & + \,  \int_{t \wedge \tau} ^{T \wedge \tau}e^{\gamma s} \, \left\|Z_s\right\|^2 ds \, + \, \int_{t \wedge \tau} ^{T \wedge \tau}e^{\gamma s} \, U_s^2 \, \lambda_s \, ds\\[10pt]
& &  - \, 2\int_{t \wedge \tau} ^{T \wedge \tau}e^{\gamma s} \, Y_s \, Z_s \, d\bar{W}_s \, - \, 2\int_{t \wedge \tau} ^{T \wedge \tau}e^{\gamma s} \, Y_{s^-} \, U_s \, dM^{}_s .\\
\end{eqnarray*}
Choosing $\gamma >1+3K+K^2$ and taking the supremum under $0$ and $T$ and the expectation, we obtain
\begin{eqnarray*}
\e \left( \, \sup_{0\leq t\leq T} e^{\gamma (t \wedge \tau)} \ Y^2_{t \wedge \tau} \right) & \leq & e^{\gamma T} \, \e \left(\xi ^2 \right) \, + \, e^{\gamma T} \, \e \left( \int_0^{T \wedge \tau}\left|f(s,0,0,0)\right|^2ds \right)\\[10pt]
 & & + \,  \e \left( \int_{0} ^{T \wedge \tau}e^{\gamma s} \, \left\|Z_s\right\|^2ds \right) \, + \, \e \left( \int_{0} ^{T \wedge \tau}e^{\gamma s} \, U_s^2 \, \lambda_s \, ds \right)\\[10pt]
&& + \, 4 \, C_{BDG} \, \e\left(\left(\int_0^{T \wedge \tau}e^{2\gamma s} \, Y^2_s \, \|Z_s\| ^2 ds\right)^{1/2} \, \right)\\[10pt]
&&+ \, 4 \, C_{BDG} \, \e \left(\left(\int_0^{T \wedge \tau}e^{2\gamma s}  \, Y^2_{s^-} \, U_s^2 \, d[M,M]_s\right)^{1/2} \, \right)\\[20pt]
&\leq & e^{\gamma T} \, \e \left(\xi ^2 \right) \, + \, e^{\gamma T} \, \e \left(\int_0^{T \wedge \tau}\left|f(s,0,0,0)\right|^2ds \right)\\[10pt]
 & & + \, \left( 1 + \frac{2}{\epsilon} \, C_{BDG} \right) \, e^{\gamma T} \, \e \left( \int_{0} ^{T \wedge \tau} \, \left\|Z_s\right\|^2ds \right) \\[10pt]
&&+ \, 4 \, \epsilon  \, C_{BDG} \, e^{\gamma T} \, \e\left(\sup_{0\leq t \leq T} \left( \, Y_{t \wedge \tau}^2 \right) \right) \\[10pt]
&& +  \, \frac{2}{\epsilon} \, C_{BDG} \, e^{\gamma T} \, \e \left(\int_0^{T \wedge \tau} U_s^2 \, d[M,M]_s\right) \, + \, e^{\gamma T}\, \e \left( \int_{0} ^{T \wedge \tau} U_s^2 \, \lambda_s \, ds \right)\, ,\\
\end{eqnarray*}
for any $\epsilon >0$.\\[20pt]
Notice that $d[M,M]_s=(\Delta H_s)^2=\Delta H_s=dH_s=dM_s+ (1-H_s) \, \lambda_s \, ds$, so applying the standard procedure of localization, one has
\[
\e \left(\int_0^{T \wedge \tau} U_s^2 \, d[M,M]_s \right) \ = \ \e \left(\int_0^{T \wedge \tau} U_s^2 \, \lambda_s \, ds \right) \, .
\]

Choosing $\epsilon \ = \ \frac{1}{8 \, C_{BDG} \, e^{\gamma T}} $, we obtain
\begin{eqnarray*}
\lefteqn{\frac{1}{2} \, \e \left(\sup_{0\leq t \leq T} \ Y_{t \wedge \tau} ^2\right) \ \leq \ e^{\gamma T} \, \e\left(\xi ^2\right) \, + \, e^{\gamma T} \, \e_{}\left(\int_0^{T \wedge \tau}\left|f(s,0,0,0)\right|^2ds \right)}\\[10pt]
&&+ \, \left( e^{\gamma T} + \frac{1}{4} \right) \, \e\left(\int_0^{T \wedge \tau}\|Z_s\|^2 ds\right) \, + \,  \left( e^{\gamma T} + \frac{1}{4} \right) \, \e_{}\left(\int_0^{T \wedge \tau} U_s^2 \, \lambda^{}_s \, ds\right)\\[20pt]
& & < \, + \infty \, .
\end{eqnarray*}

\end{document}